\documentclass[%
superscriptaddress,
twocolumn,
showpacs,
amsmath,amssymb,
aps,
prl,
floatfix,
showkeys,
reprint,
]{revtex4-1}

\usepackage{graphicx}
\usepackage{dcolumn}
\usepackage{bm}
\usepackage{amsmath}
\usepackage{amssymb}
\usepackage{multirow}
\usepackage{color}
\usepackage{booktabs}
\usepackage{tabularx}
\usepackage[english]{babel}
\usepackage[utf8]{inputenc}
\usepackage{enumerate}
\usepackage[cmyk,dvipsnames]{xcolor}
\definecolor{pacificb}{HTML}{1CA9C9}

\usepackage[normalem]{ulem}

\newcommand{\hcpIr}{hcp-Pd/Fe/Ir(111) }

\begin{document}

\title{Skyrmion lifetimes in ultrathin films}

\author{Stephan von Malottki}
\email[Corresponding author: ]{malottki@physik.uni-kiel.de}
\affiliation{Institute of Theoretical Physics and Astrophysics, University of Kiel, Leibnizstrasse 15, 24098 Kiel, Germany}

\author{Pavel F. Bessarab}
\affiliation{Institute of Theoretical Physics and Astrophysics, University of Kiel, Leibnizstrasse 15, 24098 Kiel, Germany}
\affiliation{Science Institute of the University of Iceland, 107 Reykjav\'ik, Iceland}
\affiliation{ITMO University, 197101 St. Petersburg, Russia}

\author{Soumyajyoti Haldar}
\affiliation{Institute of Theoretical Physics and Astrophysics, University of Kiel, Leibnizstrasse 15, 24098 Kiel, Germany}

\author{Anna Delin}
\affiliation{Department of Applied Physics, KTH, Kista, Sweden}
\affiliation{Department of Physics and Astronomy, Uppsala University, Sweden}
\affiliation{Swedish e-Science Research Center (SeRC), KTH Royal Institute of Technology, SE-10044 Stockholm, Sweden}

\author{Stefan Heinze}
\affiliation{Institute of Theoretical Physics and Astrophysics, University of Kiel, Leibnizstrasse 15, 24098 Kiel, Germany}

\date{\today}

\begin{abstract}
We show that thermal stability of magnetic skyrmions can be strongly affected by entropic effects.
The lifetimes of isolated skyrmions in atomic Pd/Fe bilayers on Ir(111) and on Rh(111) are calculated in the framework of
harmonic transition state theory based on an atomistic spin model parametrized from density functional theory.
Depending on the system the 
attempt frequency for skyrmion collapse can change by up to nine orders of magnitude with the strength of the applied magnetic field.
We demonstrate that this effect is due to a drastic change of entropy with skyrmion radius which opens 
a novel route towards stabilizing sub-10 nm skyrmions at room temperature.
\end{abstract}

\maketitle
 
Magnetic skyrmions are praised to be information-carrying bits in next-generation data storage and logic devices. The challenge is to obtain long-lived skyrmions at room temperature with a diameter below 10 nm -- only then emerging technologies based on skyrmions can compete with conventional solutions~\cite{fert2017magnetic,fert2013skyrmions,zhang2015magnetic}. 
While room temperature stability has been reported for skyrmions larger than 30 nm in transition-metal multilayers \cite{moreau2016additive,soumyanarayanan2017tunable}, sub-10 nm skyrmions have so far been only found in ultrathin film systems at very low temperatures of about 8~K \cite{heinze2011spontaneous,romming2013writing,romming2015field,hsu2017electric}.  
Therefore, an understanding of the mechanisms stabilizing skyrmions and ways to 
increase their lifetime at ambient temperature 
are key tasks
on the route towards skyrmion-based applications.

The skyrmion lifetime $\tau$ as a function of temperature $T$ can conveniently be described in terms of the collapse energy barrier $\Delta E$ and the attempt frequency $f_0$,
\begin{equation}
\tau=f_0^{-1} \exp{\left(\frac{\Delta E}{k_{\rm B} T}\right)},
\label{Eq:Arrhenius}
\end{equation}
with $k_B$ being the Boltzmann constant. 
The strong, exponential dependence in Eq.~(\ref{Eq:Arrhenius}) suggests that control over the energy barrier should be the key strategy for improving on thermal stability of skyrmions. 
Recent 
theoretical works have indeed focused on obtaining the energy barrier of skyrmion annihiliation~\cite{bessarab2018lifetime, vonmalottki2017enhanced, lobanov2016mechanism, uzdin2017effect, rohart2016path, bessarab2017comment, rohart2017reply,stosic_2017,cortes2017thermal,hagemeister2018controlled,Varentsova2018interplay}. 

For decades, tuning the energy barrier has been the main approach to the development of information storage bits based on magnetic materials. A rule of thumb that the energy barrier should exceed thermal energy by a factor of 40--50 at room temperature to ensure reliable information storage has become an accepted norm
(see e.g.~\cite{buettner2018,caretta2018fast}). It is important to realize, however, that such estimates are based on the assumption that the attempt frequency is a constant associated with the Larmor precession, with $f_0\approx 10^{10}$-$10^{12}$~Hz. Exclusive focus on the energy barrier can lead to biased conclusions about thermal stability when applied to unique objects such as magnetic skyrmions.  
Some hints for this claim are already present in the literature. Based on Monte-Carlo simulations, Hagemeister {\it et al.}~\cite{hagemeister2015stability} concluded that the attempt frequency for skyrmion nucleation was orders of magnitude larger than that for skyrmion annihilation. Recent theoretical works~\cite{bessarab2018lifetime,desplat2018thermal} employed statistical approaches to demonstrate that the value of the attempt frequency depends strongly on the mechanism of skyrmion collapse. Experiments by Wild and co-workers~\cite{wild2017entropy} showed that 
small variation of magnetic field 
leads to a 
dramatic
change in the attempt frequency for annihilation of skyrmions in a B20 compound 
which was interpreted as an enthalpy-entropy compensation effect. 

Here, we demonstrate by means of transition state theory (TST) and {\it ab initio} electronic structure calculations that sharp sensitivity of the attempt frequency to control parameters such as external magnetic field plays a decisive role in stabilization of nanoscale skyrmions in ultrathin films, exceeding the contribution of the energy barrier. We focus on the collapse of isolated skyrmions in the ferromagnetic background in 
%
%
%
%
%
%
%
Pd/Fe atomic bilayers on the (111) surfaces of Ir and Rh. 
The first system has been extensively studied experimentally \cite{romming2013writing,romming2015field,hagemeister2015stability,kubetzka2017impact,hanneken2015} as well as theoretically 
\cite{dupe2014tailoring,vonmalottki2017enhanced,bottcher2018b,rozsa2016complex,rozsa2017formation,bessarab2018lifetime}, while in the latter skyrmions have been recently predicted \cite{haldar2018first}. 
We find a pronounced response of the attempt frequency to the applied magnetic field close to the asymptotic divergence of the skyrmion size. We explain this behavior in terms of entropy associated with localized skyrmion modes. The entropy term rises linearly with the skyrmion surface area leading to large lifetime variations for skyrmions with diameters between 2 and 10~nm. We conclude that exploitation of entropic effects provides an efficient way to control the stability of nanoscale skyrmions.


Within the harmonic approximation to TST, the attempt frequency 
$f_0$ for isotropic skyrmion collapse is given by \cite{bessarab2018lifetime}:

\begin{equation}
f_0=\frac{2}{V_0} k_{\rm B} T \sqrt{\sum_j \frac{a_j^2}{\epsilon_{{\rm sp},j}}} \frac{\prod_i \sqrt{\epsilon_{{\rm sk},i}}}{\prod_i^{\prime} \sqrt{\epsilon_{{\rm sp},i}}}
\label{Eq:prefactor_formula}
\end{equation}
where $\epsilon_{{\rm sk},i}$ and $\epsilon_{{\rm sp},i}$ are the eigenvalues of the Hessian matrix at the skyrmion state and 
at the transition state defined by 
the saddle point (SP) on the energy surface, respectively, and $a_j$ are the components of the velocity normal to the dividing surface. 
The prime indicates that the negative eigenvalue at the SP is excluded from the product. An SP configuration corresponds to a Bloch point-like defect centered at an interstitial site of the hexagonal lattice. Therefore, there are two SPs per unit cell of the system, which results in the factor of 2 in Eq.~(\ref{Eq:prefactor_formula}). 
The energy surface as a function of the spin configuration is obtained from 
an atomistic spin model on a hexagonal lattice which contains the exchange interactions, the Dzyaloshinskii-Moriya interaction as well as the magnetocrystalline 
anisotropy energy. All interaction parameters were calculated based on {\it ab initio} density functional theory (DFT) \cite{vonmalottki2017enhanced,haldar2018first}. 
SPs on the energy surface have been obtained using the geodesic nudged elastic band (GNEB) method \cite{bessarab2015method}. 
Skyrmion translation costs almost no energy, which is reflected by a very small eigenvalue, $\epsilon_{tr}V_0\ll k_BT$, with $V_0$ being the volume of translational modes per unit cell. As a result, the translational modes are treated as Goldstone modes, leading to the linear temperature dependence of the attempt frequency \cite{bessarab2018lifetime,haldar2018first}.
\newline
\begin{figure}[h]
	\centering
  \includegraphics[width=0.47\textwidth]{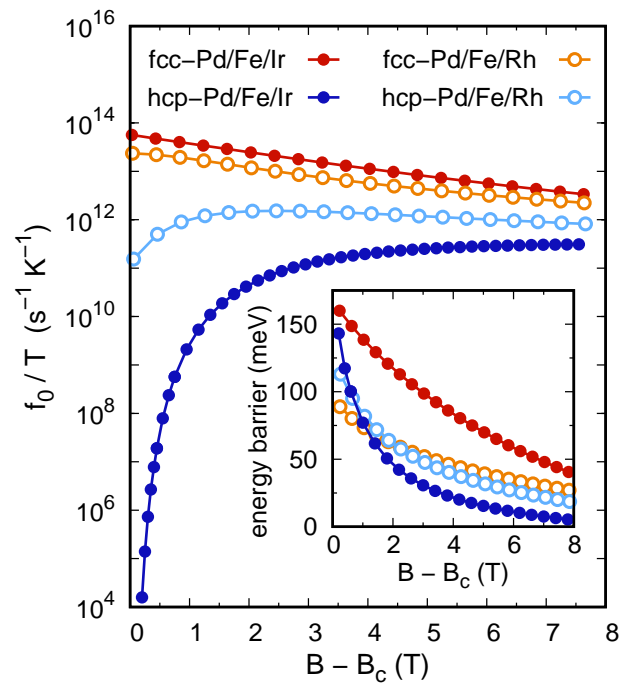}
    \caption{Calculated attempt frequencies over temperature, $f_0/T$, for fcc-Pd/Fe and hcp-Pd/Fe bilayers on Ir(111) (filled red and blue circles)
    and Rh(111) (open orange and light blue circles) 
    shown on a logarithmic scale over magnetic field strength. The critical field $B_{\rm c}$ is defined as the transition field to the ferromagnetic phase at zero temperature. $f_0$ has been calculated with HTST based on DFT parameters for the magnetic interactions. 
    The energy barriers for skyrmion collapse into the ferromagnetic state are presented in the inset.}
    \label{Fig:prefactor_barrier}
\end{figure}

The attempt frequencies were calculated based on Eq.~(\ref{Eq:prefactor_formula}) as a function of applied magnetic field for magnetic skyrmions in
Pd/Fe bilayers on Ir(111) and Rh(111) as shown in Fig.~\ref{Fig:prefactor_barrier}. Due to the linear dependence on temperature it is convenient 
to plot $f_0/T$. While the Fe layer is in fcc stacking we consider both hcp and fcc stacking of the Pd overlayer as observed in experiments \cite{kubetzka2017impact}. 
To compare all systems, the zero of the magnetic field is taken to be the critical field $B_{\rm c}$ defined as the transition field to the ferromagnetic phase
at zero temperature (see supplemental material, Table~S1).

For fcc Pd stacking, the attempt frequency decreases by roughly one order of magnitude over the considered field range for both systems.
Therefore skyrmions are stabilized by the attempt frequency for increasing fields. As a rough approximation one might still be able to use a constant
value of $f_0$ in this case.
A qualitatively different behavior is observed for hcp-Pd/Fe/Ir(111), 
where the attempt frequency increases by more than seven orders of magnitude for a magnetic field change of 
1~T.
For hcp-Pd/Fe/Rh(111), the effect is also present at small fields but there is only an increase by one order of magnitude. 
As a result of the much smaller attempt frequency the skyrmion lifetime is drastically enhanced in ultrathin 
Fe films with an hcp stacking of the Pd overlayer \footnote{We have also performed calculations in the effective, nearest-neighbor exchange approximation (cf.~Ref.~\cite{vonmalottki2017enhanced}). The change of the attempt frequency 
by orders of magnitude occurs also in these calculations.}.

\begin{figure}[h]
   	\centering
   \includegraphics[width=0.47\textwidth]{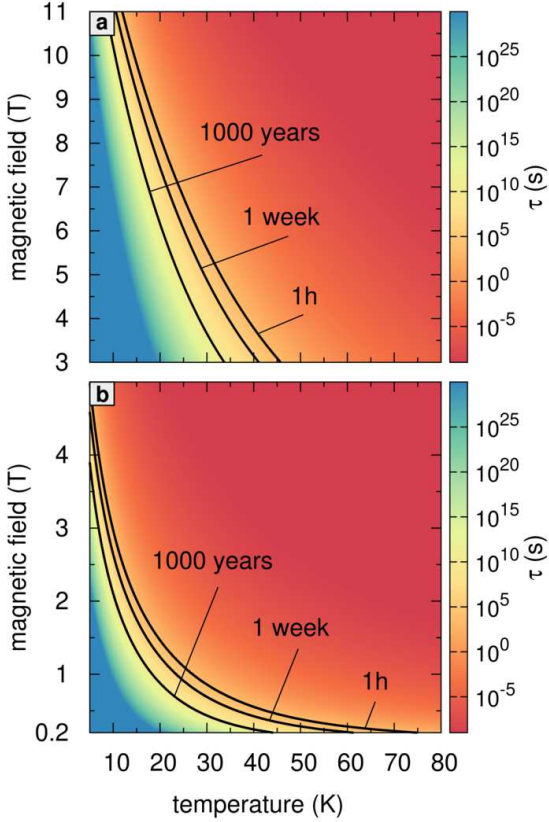}
   \caption{Mean skyrmion lifetimes for (a) fcc-Pd/Fe/Ir(111) and (b) hcp-Pd/Fe/Ir(111) are displayed by a color map for varying magnetic field strength and temperature. It is calculated by Eq.~(\ref{Eq:Arrhenius}), using data points every 0.05~T and interpolations between these points. For illustration, isolines are given for certain values of mean skyrmion lifetime.
    }
	\label{Fig:lifetimes}
\end{figure}

We combine calculated attempt frequencies with the collapse energy barriers obtained in previous works~\cite{vonmalottki2017enhanced,haldar2018first} using the GNEB method (see the inset of Fig.~\ref{Fig:prefactor_barrier}) in order to access the skyrmion lifetime via Eq.~(\ref{Eq:Arrhenius}). Calculated lifetime of skyrmions in Pd/Fe bilayers on Ir(111) is plotted in Fig.~\ref{Fig:lifetimes} as a function of the applied field and temperature 
\footnote{Similar plots for Pd/Fe/Rh(111) are shown in Ref.~\cite{haldar2018first}.}.
For both Pd stackings, the highest lifetimes are obtained for small magnetic fields and small temperatures. But while the temperature with a one hour mean lifetime amounts to $T=45$~K at the lowest accessable field ($B = 3.2$~T) for the fcc stacking of Pd, this value increases to $T=75$~K for the hcp stacking at $B = 0.2$~T. 
On the other hand, at $T=100$~K the lifetime at these magnetic field values amounts to 40~ns for fcc Pd stacking while it is 10~s for hcp Pd stacking, i.e. nine orders of 
magnitude larger.
Note that the energy barrier is even lower in the case of hcp Pd than for fcc Pd (inset of Fig.~\ref{Fig:prefactor_barrier})
and this dramatic effect is solely due to the lower attempt frequency in the hcp case which counteracts the higher energy barrier of the fcc case.

\begin{figure}[h]
	\centering
  \includegraphics[width=0.46\textwidth]{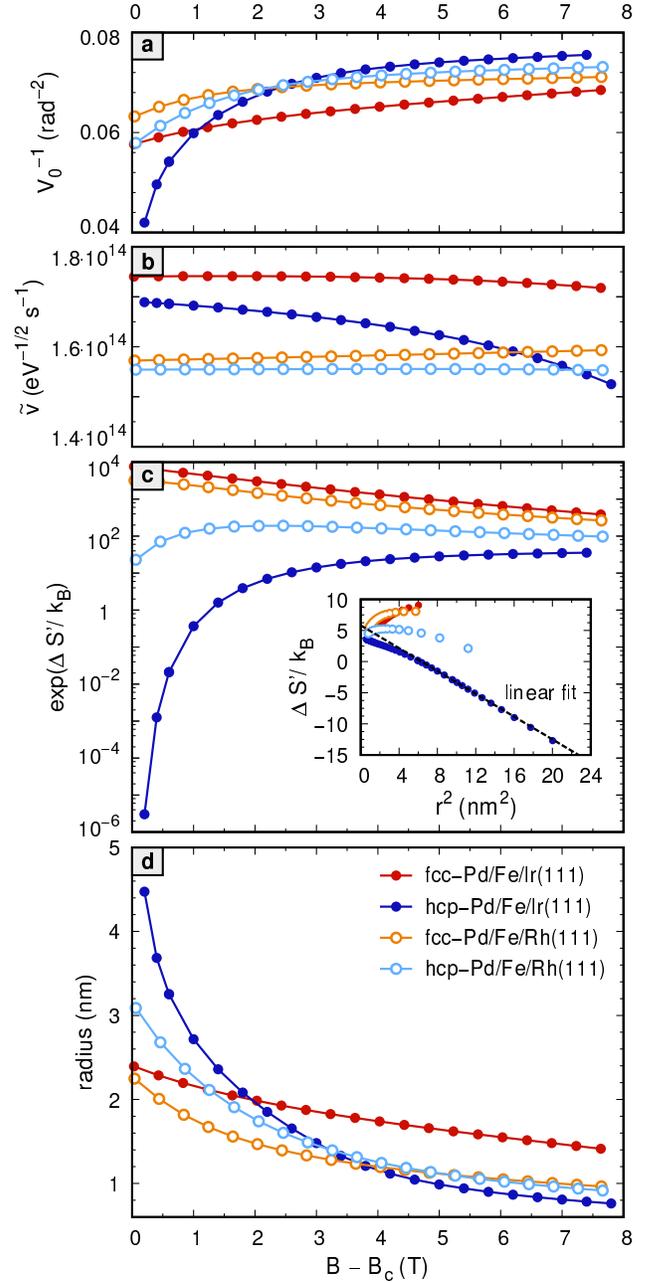}
   \caption{Terms entering the attempt frequency $f_0$ as given by
   Eq.~(\ref{Eq:prefactor_formula}) are shown separately as a function of magnetic field strength for Pd/Fe bilayers on Ir(111) and on Rh(111). 
   The inverse of the Goldstone mode volume (a), the dynamical factor $\tilde{v}$ (b) and the entropic factor $\exp \left( \Delta S^{\prime}/k_\mathrm{B} \right)$ (c) computed at $T=100$ K. In the inset of (c) $\Delta S^\prime$ is shown over $r^2$ and a linear fit has been performed 
  for \hcpIr between 12 and 24 
  nm$^2$. The radius is shown in (d).}
	\label{Fig:comp_details}
\end{figure}

The question arises why the attempt frequency displays such a strikingly different behavior for seemingly similar ultrathin film systems. To elucidate the
origin of this effect we study the system-dependent terms entering Eq.~(\ref{Eq:prefactor_formula}) separately as shown in Fig.~\ref{Fig:comp_details}. 
The inverse of the Goldstone 
mode 
volume 
$V_0$ 
[Fig.~\ref{Fig:comp_details}(a)] is monotonously increasing for all four systems and can therefore not be responsible for the different trends. Additionally, the largest increase by a factor of about two takes place in hcp-Pd/Fe/Ir(111) and is much smaller than the discussed effect. The dynamical factor $\tilde{v}\equiv\sqrt{\sum_j {a_j^2}/\epsilon_{{\rm sp},j}}$ [Fig.~\ref{Fig:comp_details}(b)] increases for Pd/Fe bilayers on Rh(111) and decreases for bilayers on Ir(111). The changes are on an even smaller scale than for the Goldstone mode volumes. Therefore, it cannot be the origin of the trend 
either. 

The third component of Eq.~(\ref{Eq:prefactor_formula}) consists of the ratio of the products of the eigenvalues. As shown in the supplemental material, it describes the contribution of the stable modes to the entropy difference between the transition state and the skyrmion state:
\begin{equation}
\sqrt{2\pi k_{\mathrm{B}} T}\frac{\prod_i \sqrt{\epsilon_{sk,i}}}{\prod_i^{\prime} \sqrt{\epsilon_{sp,i}}} = \exp \left( \frac{\Delta S^{\prime}}{k_\mathrm{B}} \right)
\label{Eq:entropy_eigenvalues}
\end{equation}
This contribution to the entropy difference is shown on a logarithmic scale in Fig.~\ref{Fig:comp_details}(c). It follows the trend of the attempt frequency and is quantitatively very similar, e.g.~it exhibits a decrease in fcc-Pd/Fe/Ir(111) by a factor of $\approx 1.5$ and an increase in hcp-Pd/Fe/Ir(111) by about seven orders of magnitude. Clearly, it is the entropy term that defines the response of the attempt frequency to the magnetic field.

\begin{figure*}
	\centering
   \includegraphics[width=0.90\textwidth]{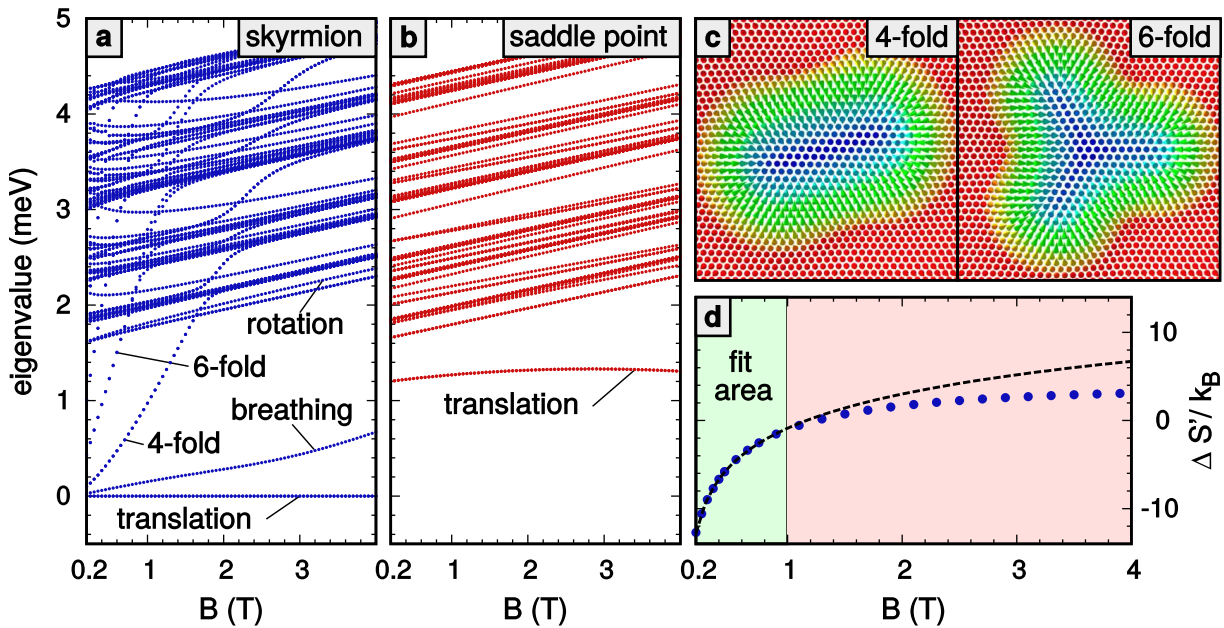}
    \caption{The calculated eigenvalues of the Hessian matrix are shown for (a) the skyrmion and (b) the saddle point of \hcpIr with varying magnetic field strengths. For the sake of clearness we only show the first 100 out of 9800 eigenvalues for both states. In (c) two skyrmion eigenvalues labeled in (a) are visualized by exciting the skyrmion along the corresponding eigenmodes. In (d) $\frac{\Delta S^{\prime}}{k_\mathrm{B}}$ is fitted by Eq.~(\ref{Eq:entropy_magnetic_field}) in the green fit area from 0.2~T to 1.0~T.}
	\label{Fig:eigenvalues}
\end{figure*}

The direct comparison of $\Delta S^\prime$ and the skyrmion radius, $r$, [Fig.~\ref{Fig:comp_details}(c) and (d)] reveals a correlation between both quantities. Note that the size of the SP  excitations are fairly constant with respect to the magnetic field (see supplemental material Fig. S1) and therefore cannot affect the trend of the entropy difference. By plotting $\frac{\Delta S^{\prime}}{k_\mathrm{B}}$ vs.~$r^2$ [inset of Fig.~\ref{Fig:comp_details}(c)] one can see that the entropy difference associated with the stable modes of hcp-Pd/Fe/Ir(111) decreases linearly with the surface area of the skyrmion for values above $r^2 \approx 4$~nm$^2$. This corresponds to magnetic fields below $\approx 2.5$~T for which the radius shows its asymptotic divergence 
with the applied field. For large fields, far away from the asymptotic behavior, the radius starts to converge to the minimal radius and the entropy difference stops following the surface area linearly. Based on this observation, the magnetic field dependence of the entropy difference and by this of the attempt frequency can be separated into an asymptotic region and a convergence region. This is consistent with both fcc systems whose radii are not asymptotic and their entropy change is not linear in $r^2$. On the other hand, the radii of hcp-Pd/Fe/Rh(111) 
suggest the beginning of an asymptotic phase in which its entropy difference starts to follow the surface area linearly.
Due to the scaling with skyrmion size we speculate that the entropy effect is even more dramatic for skyrmions in three dimensional
systems and could explain the experimental observations by Wild {\it et al.}~\cite{wild2017entropy}.

To investigate the dependencies of the entropy in more detail, we present the first 100 eigenvalues in Fig.~\ref{Fig:eigenvalues} for the skyrmion (SK) and the SP over magnetic field. 
Most of them are slightly increasing linearly and correspond to magnon or mixed modes in which the magnon modes are excited together with the SK and SP states, respectively. In Eq.~(\ref {Eq:entropy_eigenvalues}) most of these contributions should cancel out since they are similar for both states. However, some SK eigenvalues in Fig.~\ref{Fig:eigenvalues}(a) are increasing with a significantly higher slope in the asymptotic region, crossing the bands of magnon and mixed modes. These eigenvalues belong to pure skyrmion deformation modes whose 
merge into the magnon continuum is in agreement with an earlier theoretical work on skyrmions in a square lattice \cite{lin2014internal}. 
Since these skyrmion modes are only showing this behaviour in the asymptotic region, their eigenvalues seem to depend on the skyrmion size. Consistently, the SP eigenvalues do not show this features since the SP size is not changing with magnetic field (see supplemental material Fig. S1).


The assumption that only the deformation modes and the breathing mode are contributing to Eq.~(\ref {Eq:entropy_eigenvalues}) yields a direct relation of the entropy difference, $\Delta S^\prime$, to the magnetic field:
\begin{equation}
\frac{\Delta S^{\prime}}{k_\mathrm{B}} = A + \frac{n}{2} \ln \left( B \right)
\label{Eq:entropy_magnetic_field}
\end{equation}
The parameter $A$ is the sum of all particular slopes of the skyrmion eigenvalues and $n$ is the number of deformation and breathing modes. 
By fitting Eq.~(\ref{Eq:entropy_magnetic_field}) to the entropy difference over magnetic field [Fig.~\ref{Fig:eigenvalues}(d)], we found a very good agreement in the asymptotic region. 
As expected, deviations occur in the convergence region. 
We obtain $n\approx 10.3$ which is in reasonable agreement with the value of 13 contributing skyrmion modes counted at $B=0.2$ T from Fig.~\ref{Fig:eigenvalues}(a) .
%
This gives evidence that the entropy and the attempt frequency of the skyrmion are predominantly defined by only a few skyrmion eigenmodes, even though several thousand eigenvalues are entering 
Eq.~(\ref{Eq:prefactor_formula}).


In conclusion, we have demonstrated that a criterion of stability based only on the energy barrier does not hold for magnetic skyrmions at transition-metal interfaces. 
Due to entropy, the attempt frequency -- normally assumed to be constant -- can change by orders of magnitude with skyrmion diameter and system
and can become the decisive factor for skyrmion lifetime. Our work opens new paths towards stabilizing sub-10 nm skyrmions at room temperature.


We gratefully acknowledge computing time at the supercomputer of the North-German Supercomputing Alliance (HLRN).
This work was funded by the European Union’s Horizon 2020 research and innovation programme 
under grant agreement No 665095 (FET-Open project MAGicSky), the Icelandic Research Fund (Grants No. 163048-053 and 184949-051), the Russian Science Foundation (Grant No. 17-72-10195) and Alexander von Humboldt Foundation.

\bibliography{references}

\end{document}